\documentclass{article}

\usepackage{epsfig}

\begin{document}

\begin{center}
{\LARGE\bf
Is the polarized light flavor sea-quark asymmetric?
\footnote[2]{Talk presented by T. Morii at the Circum-Pan-Pacific
RIKEN Symposium on ``High Energy Spin Physics'' (RIKEN, Wako, Japan,
Nov. 3-6, 1999).}
}
\\
\vspace{1cm}
Toshiyuki {\sc Morii}
\footnote[3]{E-mail address: morii@kobe-u.ac.jp} 
and Teruya {\sc Yamanishi}$^{*,}$
\footnote[4]{E-mail address: 
yamanisi@rcnp.osaka-u.ac.jp}
\\
\vspace{1cm}
Faculty of Human Development, Kobe University, Nada, Kobe 657-8501, Japan
\\
$^*$Department of Management Sciences, Fukui University of Technology, 
Gakuen, Fukui 910-8505, Japan
\\
\vspace{1cm}
{\bf abstract}
\end{center}

Simple formulas on $\Delta\bar d(x)-\Delta\bar u(x)$
are proposed for extracting their values from polarized deep-inelastic
semi-inclusive data.  The present SMC and HERMES data suggest that
$\Delta\bar d(x)-\Delta\bar u(x)$ is slightly negative, though
the precision of the data is not enough for confirming it.

\baselineskip 20pt

\section{Introduction}

Since the NMC experiment in 1991\cite{NMC},
which precisely measured
the structure functions of the proton and neutron, $F_2^p(x)$ and $F_2^n(x)$,
for a wide region of Bjorken's $x$, it has been realized
that the Gottfried sum rule\cite{Gottfried} is violeted and
the light sea-quark distributions, $\bar u(x)$ and $\bar d(x)$,
are asymmetric.  A considerable excess of the $\bar d$ quark density
relative to the $\bar u$ quark density was seen.
The same result was confirmed by the E866 experiment which measured
the cross section ratio of the Drell-Yang processes,
$\sigma(p+d)/\sigma(p+p)$, though the
violation of the Gottfried sum rule is smaller than reported by the NMC.
Now, study on the origin of the flavor asymmetry of light
sea--quarks has been a challenging subject in particle and nuclear physics
because it is closely related to the dynamics of nonperturbative
QCD\cite{Kumano}.
Although the most widely accepted idea to understand it is
the meson cloud model, many discussions are still under going with
several approaches such as chiral quark model, Skyrme
model, Pauli blocking effects, etc.\cite{Steffens}.

However, what is going on with the polarized case, $\Delta\bar u(x)$
and $\Delta\bar d(x)$?
In these years, measurement of the polarized structure function of the
nucleon in polarized deep--inelastic scatterings have shown
that the nucleon spin is carried by quarks
a little and the strange sea--quark is negatively polarized in quite large.
The results were not anticipated by conventional theories
and often referred as `{\it the proton spin
crisis}'\cite{Lampe}.
By using many data with high precision on the
polarized structure functions of the proton, neutron and
deuteron accumulated so far, good parametrization models of
polarized parton distribution functions have been proposed
at the next--to--leading order(NLO) of QCD\cite{polPDF}.
The behavior of polarized valence $u$ and $d$ quarks has been well--known
from such analyses.  However, the knowledge of polarized sea--quarks
and gluons is still poor.  Although people usually assume the
symmetric light sea--quark polarized distribution, i.e.
$\Delta\bar u(x)=\Delta\bar d(x)$, in analyzing the polarized
structure functions of nucleons, there is no physical ground of such an
assumption.  In order to understand the nucleon spin structure,
it is very important to know if the light sea--quark flavor symmetry
is broken even for polarized distributions and to determine
how $\Delta\bar u(x)$ and $\Delta\bar d(x)$ behave in the nucleon.
Related to these subjects, it is interesting to know
that even if we start with the symmetric distributions for the polarized
light sea--quarks, $\Delta\bar u=\Delta\bar d$, at an initial
$Q_0^2$, the symmetry can be violated for higher $Q^2$ regions, if
the polarized distributions are perturbatively evolved in NLO
calculations of QCD\cite{Stratmann}.
In addition, some people have estimated the amount of its
violation at an initial $Q^2_0$ using some effective models.
However, their results do not agree with each other\cite{Fries,Dressler}.
Therefore, it is interesting to extract the value of
$\Delta\bar d(x)-\Delta\bar u(x)$ from the experimental data and
test the flavor symmetry of $\Delta\bar d(x)$ and $\Delta\bar u(x)$
experimentally.

Recently, using longitudinal polarized lepton beams and longitudinal
polarized fixed targets,  SMC group at
CERN\cite{SMC98} and HERMES group at DESY\cite{Simon,Funk}
observed the cross sections of the following semi--inclusive processes,
\begin{equation}
\vec l + \vec N \to l' + h + X~,
\label{eqn:SI}
\end{equation}
and obtained the data on spin asymmetries for proton, deuteron and $^3$He
targets, where $h$ is a created charged hadron or one of  $\pi^{\pm}$,
$K^{\pm}$, $p$ and $\bar p$.
A created hadron depends on the flavor of a
parent quark and thus properly combining these data it is possible to
decompose polarized quark
distributions into the ones with individual flavor\cite{Frankfurt}.
These data provide a good material to test the light flavor symmetry
of polarized sea--quark distributions and it might be timely to test
the symmetry by using the present data.

\section{Simple formulas of $\Delta\bar d(x)-\Delta\bar u(x)$}

In this letter, we propose new formulas for extracting
a difference, $\Delta\bar d - \Delta\bar u$, from the data
of the above--mentioned semi--inclusive processes and estimate the
value of it from the
present data in order to test if the light flavor symmetry of
polarized sea--quark distributions is originally violated.

Let us start with the semi--inclusive asymmetry for the process
of eq.(\ref{eqn:SI})
with proton targets, which is written by\cite{SMC98}
\begin{equation}
A_{1 p}^h(x, Q^2)~=
\frac{\sum_{q, H}~e_q^2~\{\Delta q(x, Q^2)~D^H_q(Q^2)+
\Delta \bar q(x, Q^2)~D^H_{\bar q}(Q^2)\}}
{\sum_{q, H}~e_q^2~\{q(x, Q^2)~D^H_q(Q^2)+\bar q(x, Q^2)~D^H_{\bar q}(Q^2)\}}~
\times \{ 1 + R(x, Q^2)\},
\label{eqn:A1}
\end{equation}
in the leading order(LO) of
QCD\cite{Altarelli}, where
$\Delta q(x, Q^2)$ ($\Delta \bar q(x, Q^2)$) and
$q(x, Q^2)$ ($\bar q(x, Q^2)$) are the spin--dependent
and spin--independent quark distribution functions
at some values of $x$ and $Q^2$, respectively, and
$R(x, Q^2)$ is a ratio of the absorption cross section of longitudinally
and transversely polarized virtual photons by the nucleon,
$R(x, Q^2)=\sigma _L/\sigma _T$.
$D^H_q(Q^2)$ is given by integration of the fragmentation function,
$D^H_q(z, Q^2)$, over the measured kinematical region of $z$, i.e.
$D^H_q(Q^2)=\int^1_{z_{min}}dz~D^H_q(z, Q^2)$, where
$D^H_q(z, Q^2)$ represents
the probability of producing a hadron $H$ carrying momentum fraction
$z$ at some $Q^2$ from a struck quark with flavor $q$.  $h$ is the
observed hadron concerned with here.
When $h$ is $h^+$, the fragmentation function of, for example, $u$--quark
decaying into $h^+$ is given by
\begin{equation}
D^{h^+}_u(z, Q^2)=D^{\pi^+}_u(z, Q^2)+D^{K^+}_u(z, Q^2)+D^p_u(z, Q^2)~,
\label{eqn:FRG-U}
\end{equation}
because $h^+$ is dominantly composed of $\pi^+$, $K^+$ and $p$.
Assuming the reflection symmetry along the
V--spin axis, the isospin symmetry and charge conjugation invariance of
the fragmentation functions, many fragmentation functions
can be classified into the following 6 functions\cite{Kumano},
\begin{eqnarray}
&&D\equiv D^{\pi^+}_u=D^{\pi^+}_{\bar d}=D^{\pi^-}_d=D^{\pi^-}_{\bar u}~~,
\nonumber\\
&&\widetilde{D}\equiv D^{\pi^+}_d=D^{\pi^+}_{\bar u}=D^{\pi^-}_u
=D^{\pi^-}_{\bar d}=D^{\pi^+}_s=D^{\pi^+}_{\bar s}
=D^{\pi^-}_s=D^{\pi^-}_{\bar s}~~,\nonumber\\
&&D^K\equiv D^{K^+}_u=D^{K^+}_{\bar s}=D^{K^-}_{\bar u}=D^{K^-}_s~~,
\label{eqn:FRA}\\
&&\widetilde{D^K}\equiv D^{K^+}_d=D^{K^+}_s=D^{K^+}_{\bar u}=D^{K^+}_{\bar d}
=D^{K^-}_u=D^{K^-}_d=D^{K^-}_{\bar d}=D^{K^-}_{\bar s}~~,\nonumber\\
&&D^p\equiv D^p_u=D^p_d=D^{\bar p}_{\bar u}=D^{\bar p}_{\bar d}~~,\nonumber\\
&&\widetilde{D^p}\equiv D^p_s=D^p_{\bar u}=D^p_{\bar d}=D^p_{\bar s}
=D^{\bar p}_u=D^{\bar p}_d=D^{\bar p}_s=D^{\bar p}_{\bar s}~~,\nonumber
\end{eqnarray}
where $D^H$ and $\widetilde{D^H}$ are called favored and unfavored
fragmentation functions, respectively.  Here we follow the commonly
taken assumption on the fragmentation functions, for simplicity.

Now, we can rewrite eq.(\ref{eqn:A1}) as
\begin{eqnarray}
& &\sum_{q, H}~e_q^2~\{\Delta q(x, Q^2)~D^H_q(Q^2)+
\Delta \bar q(x, Q^2)~D^H_{\bar q}(Q^2)\}\nonumber\\
& &~~~~~~~~~~~~=
\frac{A_{1 p}^h(x, Q^2)~[\sum_{q, H}~e_q^2~
\{q(x, Q^2)~D^H_q(Q^2)+\bar q(x, Q^2)~D^H_{\bar q}(Q^2)\}]}
{\{ 1 + R(x, Q^2)\}}~\nonumber\\
& &~~~~~~~~~~~~= \Delta N^h_p(x, Q^2)~,
\label{eqn:Dq}
\end{eqnarray}
where $\Delta N^h_p(x, Q^2)$ is reffered to the spin--dependent
production processes of charged hadrons with proton targets. From
a combination of $\Delta N^{h^+, h^-}_{p, n}(x, Q^2)$ for proton and
neutron targets, we can obtain the following formula,
\begin{eqnarray}
&&\hspace{-0.5cm}\Delta\bar d(x, Q^2)-\Delta\bar u(x, Q^2)\nonumber\\
&&\hspace{0.5cm}=\frac{\Delta N^{h^+}_p(x, Q^2)-\Delta N^{h^+}_n(x, Q^2)-
\Delta N^{h^-}_p(x, Q^2)+\Delta N^{h^-}_n(x, Q^2)}{2~I_1(Q^2)}\nonumber\\
&&\hspace{0.5cm}-\frac{\Delta N^{h^+}_p(x, Q^2)-\Delta N^{h^+}_n(x, Q^2)+
\Delta N^{h^-}_p(x, Q^2)-\Delta N^{h^-}_n(x, Q^2)}{2~I_2(Q^2)}~,
\label{eqn:hadron}
\end{eqnarray}
where
\begin{eqnarray}
&&I_1(Q^2)=5D(Q^2)+4D^K(Q^2)+3D^p(Q^2)-
5\widetilde{D}(Q^2)-4\widetilde{D^K}(Q^2)-3\widetilde{D^p}(Q^2)~,\nonumber\\
&&I_2(Q^2)=3D(Q^2)+4D^K(Q^2)+3D^p(Q^2)+
3\widetilde{D}(Q^2)+2\widetilde{D^K}(Q^2)+3\widetilde{D^p}(Q^2)~.\nonumber\\
\end{eqnarray}
Furthermore, if one can specify the detected charged hadron in experiment,
one can obtain more simplified formulas for the difference of
polarized light sea--quark densities.
For the case of semi--inclusive $\pi^{\pm}$--productions with proton and
neutron targets, the difference can be written by
\begin{eqnarray}
&&\hspace{-0.5cm}\Delta\bar d(x, Q^2)-\Delta\bar u(x, Q^2) \nonumber\\
&&=\frac{1}{6\{D(Q^2)+\widetilde{D}(Q^2)\}}
\times [\{J(Q^2)-1\}\{\Delta N^{\pi^+}_p(x, Q^2)-
\Delta N^{\pi^+}_n(x, Q^2)\} 
\label{eqn:pi} \\
&&\hspace{4cm}-
\{J(Q^2)+1\}\{\Delta N^{\pi^-}_p(x, Q^2)-\Delta N^{\pi^-}_n(x, Q^2)\}]~,
\nonumber
\end{eqnarray}
where
$J(Q^2)=\frac{3(D(Q^2)+\widetilde{D}(Q^2))}{5(D(Q^2)-\widetilde{D}(Q^2))}$.
Eqs.(\ref{eqn:hadron}) and (\ref{eqn:pi}) are main results of this work.
Based on these formulas, one can extract
$\Delta\bar d(x, Q^2) - \Delta\bar u(x, Q^2)$ by using the
values of $\Delta N^h_N(x, Q^2)$ which can be derived
from experimental data of spin asymmetries
$A_{1 N}^h(x, Q^2)$, if the
spin--independent quark distribution functions and
fragmentation functions are well known.

\section{Extraction of $\Delta\bar d(x)-\Delta\bar u(x)$ from 
semi-inclusive data}

The remaining task is to numerically estimate the value of
$\Delta\bar d(x, Q^2) - \Delta\bar u(x, Q^2)$ from the present
semi--inclusive data in order to examine how these formulas
are effective for testing the light flavor symmetry of polarized
distributions.
In this analysis, we use the parametrization of GRV98(LO)\cite{GRV98}
for the unpolarized parton distribution being the $\bar u/\bar d$
asymmetric and the R1990 parametrization\cite{R1990} for the ratio $R$
in eq.(\ref{eqn:A1}).
The fragmentation functions of eq.(\ref{eqn:FRA}) are determined
so as to fit well the EMC data\cite{EMC} and by
integrating them from $z_{min}=0.2$ to $1$, we have obtained
$D^H(Q^2)$ and $\widetilde D^H(Q^2)$.
At present, we have some data of $A_{1 p}^{h^{\pm}}$ and $A_{1 d}^{h^{\pm}}$
measured by the SMC group and also some data of
$A_{1 p}^{h^{\pm}}$, $A_{1 ^3He}^{h^{\pm}}$,
$A_{1 p}^{\pi^{\pm}}$ and $A_{1 ^3He}^{\pi^{\pm}}$
by the HERMES group. From
these data, we can estimate the values of
$\Delta\bar d(x, Q^2) - \Delta\bar u(x, Q^2)$
from eqs.(\ref{eqn:hadron}) and (\ref{eqn:pi}) by using
$\Delta N_N^h$ calculated from the data set of ($A_{1 p}^{h^{\pm}}$,
$A_{1 d}^{h^{\pm}}$) by SMC and ($A_{1 p}^{h^{\pm}}$, $A_{1 ^3He}^{h^{\pm}}$)
and ($A_{1 p}^{\pi^{\pm}}$, $A_{1 ^3He}^{\pi^{\pm}}$)
by HERMES.
Here, for the data of $^3$He targets, the values of $A_{1n}^h$ were derived
from the data of $A_{1 ^3He}^h$ according to the way in ref.\cite{Funk}.
In the present analysis, we have neglected the $Q^2$ dependence
being fixed as $Q_0^2=4$GeV$^2$
because no significant $Q^2$ dependence has been observed in this region
in the spin asymmetry $A_{1 N}$ for inclusive data\cite{SMC97}.
The results calculated from eqs.(\ref{eqn:hadron}) and (\ref{eqn:pi})
are presented in fig.1.
We have checked the model dependence of unpolarized
quark distribution functions and found
that the results are not sensitive to
those models.

\begin{figure}[h]
\center
\epsfig{file=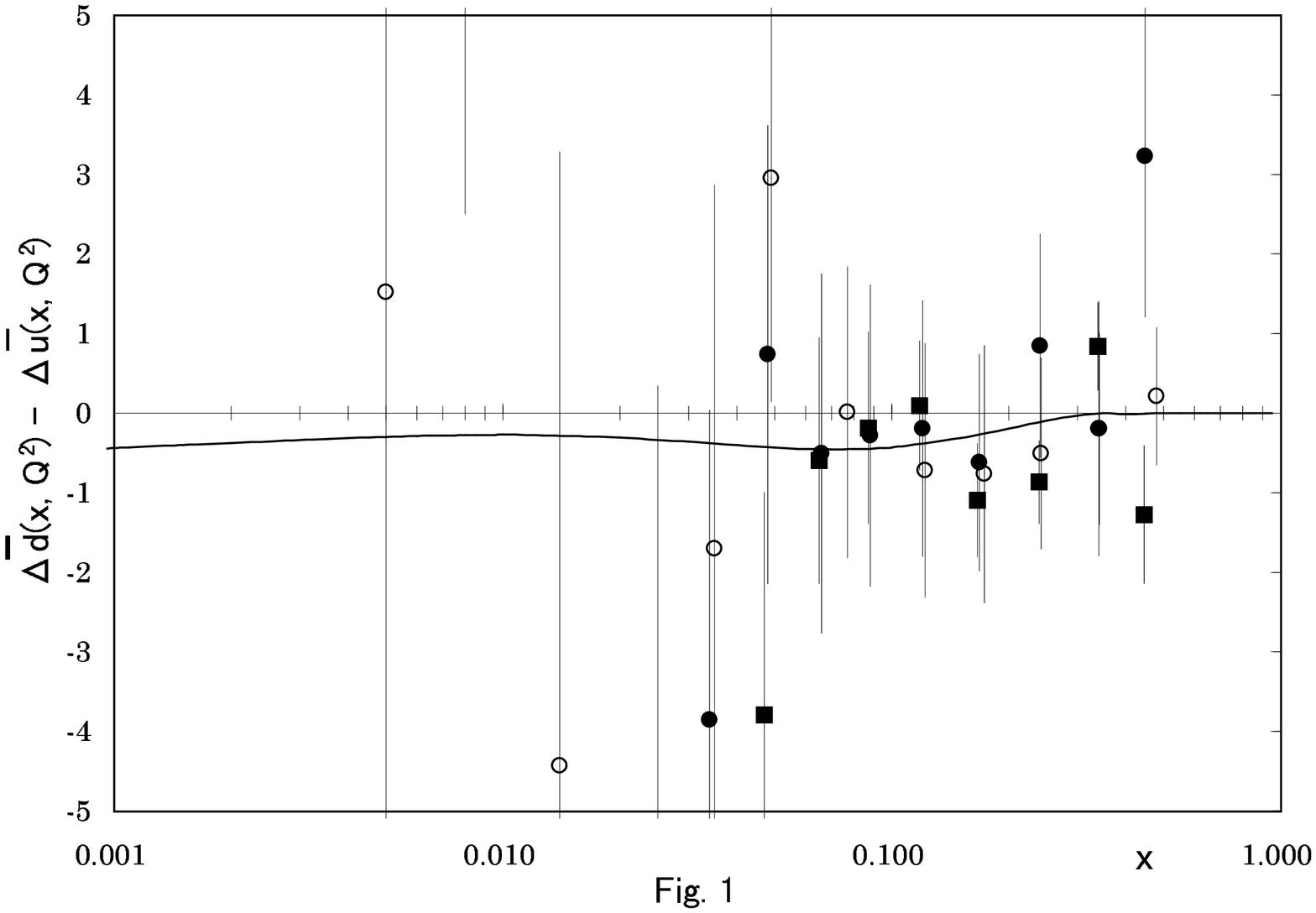,height=12cm,width=7.5cm,angle=-90}
\caption{
The $x$ dependence of $\Delta\bar d(x, Q^2)-\Delta\bar u(x, Q^2)$ 
at $Q^2=4\ {\rm GeV}^2$ estimated by using the GRV98(LO) 
and R1990 parametrizations for the unpolarized quark distributions 
and ratio R, respectively.
Marks indicated by the solid circle, open circle and solid square denote 
the results calculated from the data set of SMC data, 
HERMES data for charged hadron productions and HERMES data 
for charged pion productions, respectively.
The solid line indicates the result of $\chi^2$-fit by 
the parametrization of eq. (9).
}
\end{figure}

To examine the behavior of $\Delta\bar d(x)-\Delta\bar u(x)$ in more detail
and to test the light flavor asymmetry of $\Delta\bar d(x)$ and
$\Delta\bar u(x)$, we have parametrized it as
\begin{equation}
\Delta\bar d(x)-\Delta\bar u(x)=Cx^{\alpha}(\bar d(x)-\bar u(x)),
\label{eqn:chifit}
\end{equation}
and determined the values of $C$ and $\alpha$ from
the $\chi ^2$--fit to the results presented in fig.1.
The results were
$C=-3.40(-3.87)$ and $\alpha =0.567(0.525)$ for the
GRV98(LO)\cite{GRV98}(MRST98(LO)\cite{MRST}) unpolarized
distributions, while the values of $\chi ^2$/{d.o.f.} were
0.91(0.90) for
GRV98(LO)(MRST98(LO)).  $C<0(\not=0)$ is a remarkable result,
suggesting an asymmetry of $\Delta\bar d(x)$ and $\Delta\bar u(x)$.
It is
interesting to note that the negative value of $C$ is consistent with
instanton interaction predictions\cite{Dorokov}.
Also, the similar result is indicated from the chiral quark
soliton model\cite{Wakamatsu}.
However, it must be premature to lay stress on this result because
of too large errors of the present data,
though this result might suggest a violation of the polarized
light flavor sea--quark symmetry.
We urge to have more data with high precision to confirm this result.

Some comments are in order for the usefulness of our formulas: (i) Our
formulas depend on the unpolarized parton distribution
functions and the fragmentation functions.  Unfortunately, some of them
are poorly known at present.  In addition, $\Delta N_{p(n)}^h$ depends
on the semi--inclusive asymmetry, $A_{1p(n)}^h$, and contains some
experimental errors.  Therefore, it might be rather difficult to
extract the exact value of $\Delta\bar d(x) - \Delta\bar u(x)$ from the
present data.  However, we believe that they must be quite useful for future
experimental test of the polarized light sea-quark asymmetry if we have
more precise data and good information on these functions.  Our formulas
are simple and can be easily tested in experiment.
(ii) At present we see only asymmetries, $A_{1p(n)}^h$, in literature.
However, if the precise experimental data on the polarized cross sections
will be presented, then our formulas make more sense by replacing
$A_{1p(n)}^h$ by the polarized cross sections themselves, where the
unpolarized parton distributions and $R(x, Q^2)$ do not come in and we do
not need to worry about their uncertainty.

\section{Summary and discussion}

In conclusion, we have proposed simple new formulas for
extracting a difference of the polarized light sea--quark density,
$\Delta\bar d(x)-\Delta\bar u(x)$, from polarized deep--inelastic
semi--inclusive data and numerically estimated it
using these formulas
from the present experimental
data for semi--inclusive processes.
Unfortunately, the precision of the present data is not enough for
extracting an exact value of the difference,
$\Delta\bar d(x)-\Delta\bar u(x)$, and unambiguously
testing the polarized light flavor sea--quark asymmetry.
However, the HERMES group is now measuring semi--inclusive processes
by using a new detector called RICH which can identify
each of charged particles over a wide kinematical range and
these data of charged pions with high statistics are expected to
allow us to
test more clearly the asymmetry of polarized light flavor sea--quark
densities.

Another interesting way to study the asymmetry of $\Delta\bar u(x)$ and
$\Delta\bar d(x)$ is the Drell--Yan process for polarized
proton/deuteron--polarized proton collisions\cite{Kumano99}.
The process provides informations on the raito of
$\Delta\bar u(x)/\Delta\bar d(x)$ and thus it
is complementary to our processes.

\section*{Acknowledgements}

One of us (T. Y.) thanks S. Kumano, H. Kitagawa and Y. Sakemi
for valuable discussions.

\end{document}